\documentclass[showpacs,prb,aps,twocolumn,fontsize=12pt,floatfix]{revtex4-1}
\usepackage{amsmath}
\usepackage[normalem]{ulem}
\usepackage{amsfonts}
\usepackage{amssymb}
\usepackage{enumitem}
\usepackage{amsthm}
\usepackage{graphicx}
\usepackage{color}
\usepackage[utf8]{inputenc}
\usepackage[verbose,hypertexnames=false,bookmarksopenlevel=1,filecolor=blue,
linkcolor=blue,citecolor=blue,pdfstartview=FitH,bookmarksopen,bookmarksnumbered,
colorlinks,plainpages=false,linktocpage]{hyperref}
\definecolor{orange}{rgb}{1,0.5,0}

\begin{document}
%----------------------------------------------------------------%
%----------------------------------------------------------------%
\title{Entanglement spectrum of the degenerative ground state of Heisenberg ladders in a time-dependent magnetic field}
\author{Sonja Predin}
\email[]
{sonja.predin@physik.uni-regensburg.de}
\affiliation{ Institute for Theoretical Physics, University of
Regensburg, D-93040 Regensburg, Germany}


\begin{abstract}
We investigate of the relationship between the entanglement and 
subsystem Hamiltonians in the perturbative regime of strong 
coupling between subsystems. 
One of the two conditions that guarantees the proportionality 
between these Hamiltonians obtained by using the nondegenerate 
perturbation theory within the first order is that the 
unperturbed ground state has a trivial entanglement 
Hamiltonian. Furthermore, we study the entanglement 
Hamiltonian of the Heisenberg ladders in a time-dependent 
magnetic field using the degenerate 
perturbation theory, where couplings between legs are 
considered as a perturbation. 
In this case, when the ground state is two-fold degenerate,
and the entanglement Hamiltonian is proportional to the 
Hamiltonian of a chain within first-order perturbation theory,
even then also the unperturbed ground state has a nontrivial 
entanglement spectrum.
\end{abstract}

\maketitle

\section{Introduction}
Quantum entanglement, primarily a source of quantum information, 
has developed into one of the most studied subfields of many-body physics.
In the last decade, quantum entanglement has mainly been used to study 
phase structure in condensed matter physics \cite{Amico2008}.
The entanglement spectrum of a bipartite system of subsystems A and B
is defined in terms of the Schmidt decomposition of its ground state
$ \vert \psi \rangle $ as
\begin{eqnarray}
\vert \psi \rangle = \sum_{n} e^{-\frac{\xi_{n}}{2}} 
\vert \psi_{n}^{A} \rangle \vert \psi_{n}^{B} \rangle
\end{eqnarray}
where the states $ \vert \psi_{n}^{A} \rangle $ ($ \vert \psi_{n}^{B} \rangle $) 
are orthonormal states of the subsystem
A (B), respectively, and the non-negative quantities $ \xi_{n} $ represent 
the levels of the entanglement spectrum. 
Further, Haldane and Li  in Ref.\cite{Li08} have reported a remarkable relationship 
between the excitation spectrum and the edges separating the subsystems, 
considering the entanglement spectrum of the fractional quantum Hall system 
obtained using a spatial cut. This connection between
the edge spectrum and entanglement spectrum is observed in many condensed matter systems including ladders systems
\cite{Qi12, Chen13, Lundgren13} 
In many previous studies, the proportionality between the energetic Hamiltonian
of the subsystem A $ \mathcal{H}_{A} $ and the entanglement Hamiltonian 
$ \mathcal{H}_{ent} $ in the strong coupling regime \cite{POI10, CIR11,
SCH11, PES11, LAE12, SCH12, SCH13} has been observed.
However, this does not hold in general, even in the strong coupling limit
which is illustrated by counterexamples in Ref. \cite{LUN12}where four spin terms of the Kugel-Khomskii model
are considered in Ref. \cite{SCH12}, in which anisotropic spin ladders of arbitrary
spin length were considered, where even the unperturbed nondegenerate ground state
has a nontrivial entanglement spectrum.

Here we study the entanglement spectrum of the Heisenberg spin 
ladders in a time-dependent magnetic field via the degenerate perturbation theory,
where couplings between legs are considered as a perturbation. 
The entanglement Hamiltonian is, within the first-order perturbation theory,
proportional to the energy Hamiltonian of a chain in the magnetic field
when the ground state is degenerate.
This holds, although the entanglement spectrum of the unperturbed
ground state has a nontrivial entanglement spectrum.

\section{Motivation}
\label{motivacija}

\begin{figure}[t]
\includegraphics[width=1\columnwidth]{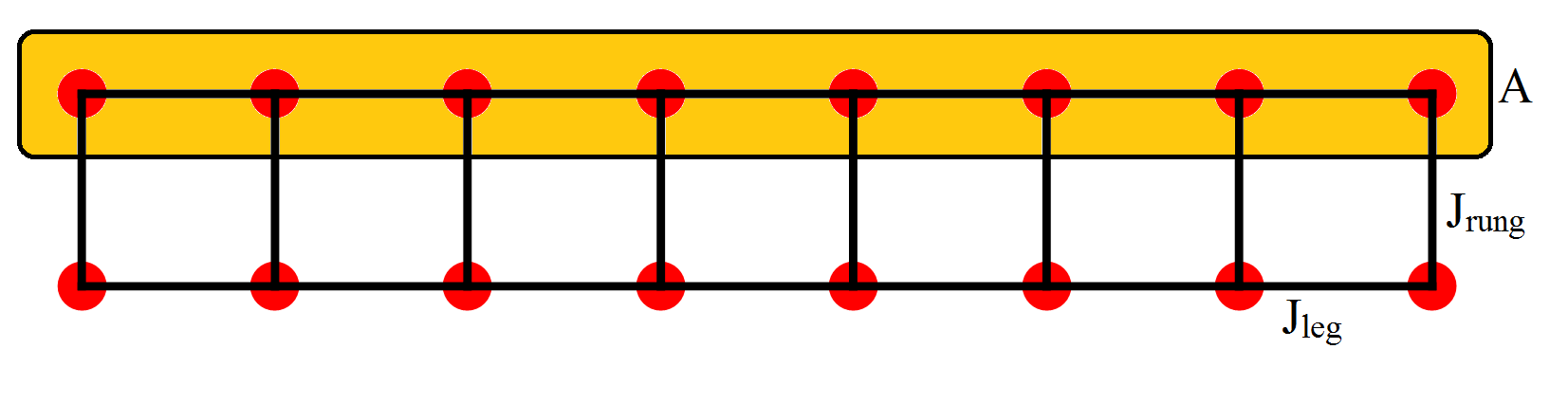}
\caption{Illustration of the two leg spin ladder considered in this paper. The entanglement spectrum is performed by tracing out subsystem A.
}
\label{fig1}
\end{figure}

We consider a bipartite system consisting of two subsystems described
by Hamiltonians $ \mathcal{H}_{A} $ of subsystem A and $ \mathcal{H}_{B} $ of subsystem B, 
which are coupled by the Hamiltonian $ \mathcal{H}_{0} $. We assume that the Hamiltonians 
$ \mathcal{H}_{A} $ and $ \mathcal{H}_{B} $ are small compared to $ \mathcal{H}_{0} $,
and it will be treated as a small perturbation.
This problem can be illustrated by two leg spin ladders, where interaction between rungs is considered as a small perturbation (see Fig. \ref{fig1}).

The projector onto subsystem orthogonal on the
nondegenerative ground state $ \vert \psi_{0} \rangle $ is defined as
\begin{equation}
Q_{l} = 1 - \vert \psi_0 \rangle \langle \psi_0 \vert = \sum_{l\neq 0} \vert \psi_l
\rangle \langle \psi_l \vert.
\end{equation}
Then, the first correlation $ \vert \psi_{l}^{(1)} \rangle $ of the nondegenerative
ground state $ \vert \psi_{0} \rangle $ reads
\begin{small}
\begin{align}
\vert \psi_{l}^{(1)} \rangle = &  \vert \psi_{0} \rangle \nonumber \\
+ &
\frac{1}{E_{0}-\mathcal{H}_{0}}Q_{l}\left(\left(\mathcal{H}_{A}+\mathcal{H}_{B}\right)-
\langle \psi_{0}\vert \left(\mathcal{H}_{A}+\mathcal{H}_{B}\right) \vert \psi_{0} \rangle \right)
\vert \psi_{l} \rangle
\label{def_corr2}
\end{align}
\end{small}
where $ E_{0} = \langle \psi_{0} \vert \mathcal{H}_{0} \vert \psi_{0} \rangle $. 
We also use that $ \frac{1}{E_{0}-\mathcal{H}_{0}} Q_{l}=
\sum_{l\neq 0}\frac{\vert\psi_{l}\rangle \langle \psi_{l}\vert}{E_{0}-E_{l}} $,
where $ E_{l} = \langle \psi_{l} \vert \mathcal{H}_{0} \vert \psi_{l} \rangle $
and the fact that $ \frac{1}{E_{0}-\mathcal{H}_{0}} Q_{l} \vert \psi_{0} \rangle = 0 $
by definition.
In the following, we will assume that $ \frac{1}{E_{0}-E_{l}} $ is equal for every $ l $.
This allows us to rewrite Eq.(\ref{def_corr2}) as
\begin{eqnarray}
\vert \psi_{l}^{(1)} \rangle = \vert \psi_{0} \rangle + \frac{1}{E_{0}-E_{l}}
 (\mathcal{H}_{A}+\mathcal{H}_{B}) \vert \psi_{0} \rangle.
\end{eqnarray}
The density matrix within the first order of the perturbation theory
has the following form
\begin{small}
\begin{align}
\rho =& \vert \psi_{l}^{(1)} \rangle \langle \psi_{l}^{(1)} \vert \nonumber \\
\rho =& \vert \psi_{0} \rangle \langle \psi_{0} \vert \nonumber
\\ &+ 
\frac{1}{E_{0}-E_{l}}
( (\mathcal{H}_{A}+\mathcal{H}_{B})\vert \psi_{0} \rangle
 \langle \psi_{0} \vert + \vert \psi_{0} \rangle \langle \psi_{0} \vert
  (\mathcal{H}_{A}+\mathcal{H}_{B})).
\end{align}
\end{small}
Owing to the fact that, here, the Hamiltonian $ \mathcal{H}_{A} $ acts only on the subsystem A,
the reduced density matrix can be calculated by tracing out the subsystem B
\begin{eqnarray}
\rho_{red}=\rho_{1} + \frac{1}{E_{0}-E_{l}}
\left(\mathcal{H}_{A} \rho_{1} + \rho_{1} \mathcal{H}_{A} \right)
\end{eqnarray} 
where $ \rho_{1} = {\textrm tr}_{2} \vert \psi_{0} \rangle \langle \psi_{0} \vert $
is the reduced density matrix within the zeroth order of the perturbation theory.
When the two subsystems are maximally entangled, $ \rho_{1} $ is proportional to the unit matrix. 
In this case, we obtain
\begin{eqnarray}
\rho_{red} = \rho_{1}\left(1-\frac{2}{E_{0}-E_{l}}\mathcal{H}_{A}\right).
\end{eqnarray}
The reduced density matrix can be reformulated as
\begin{eqnarray}
\rho_{red}=\frac{1}{Z}\exp(-\mathcal{H}_{ent}^{(1)})
\end{eqnarray}
where the entanglement Hamiltonian $ \mathcal{H}_{ent} $ is the entanglement
Hamiltonian, and the partition function $Z$ is $ Z=tr\left(exp(-\mathcal{H}_{ent}^{(1)})\right) $.
The entanglement Hamiltonian
\begin{eqnarray}
\mathcal{H}_{ent}=\frac{2}{E_{0}-E_{l}}\mathcal{H}_{A}
\end{eqnarray}
is proportional to the Hamiltonian of subsystem A, with the proportionality
factor $ \beta = \frac{2}{E_{0}-E_{l}} $ interpreted as an inverse temperature.

To conclude, we assume that
\begin{enumerate}[label=\textnormal{(\arabic*)}]
\item $ \frac{1}{E_{0}-E_{l}} $ is equal for every $ l $, and\label{itm:1}
\item $ \rho_{1} $ is proportional to the unit matrix\label{itm:2}.
\end{enumerate}
These two assumptions directly lead to the proportionality between
the entanglement and subsystem Hamiltonians in the strong coupling limit
within the first order of the perturbation theory, when the ground state is
nondegenerate.

In Ref. \cite{POI10}, Poilblanc stressed a remarkable similarity between the chain--
chain entanglement spectrum in the two-leg spin-1/2 ladders and the energy spectrum of a 
single spin-1/2 Heisenberg chain.
%Furthermore, he found that the effective inverse temperature depends on the ration 
%of the leg to the rung couplings and vanishes in the limit of the strong rung coupling.
L\"auchli and Schliemann \cite{LAE12} analytically showed that the entanglement 
Hamiltonian of the two coupled anisotropic XXZ chains is proportional to the energy 
Hamiltonian of the single chain with renormalized anisotropy in the first order of the 
perturbation theory in the strong coupling limit. There, the first assumption \ref{itm:1}
is not valid. In the case of the 
isotropic Heisenberg ladders, both assumptions \ref{itm:1} and \ref{itm:2} are valid, and for that reason,
they found that the entanglement spectrum is directly 
proportional to the energy of the single chain. 
%The proportionality factor, the inverse temperature, depends on the ratio of the leg to 
%the rung couplings.
The authors in Ref. \cite{SCH12} generalized this observation 
for the isotropic Heisenberg ladders to the case of the arbitrary spin length 
$ \textit{S} $. They found that for arbitrary spin, the entanglement spectrum 
of the isotropic Heisenberg ladders is proportional to the energy of the single chain 
within the first-order perturbation theory. This is also a consequence of the fact that both
assumptions \ref{itm:1} and \ref{itm:2} hold. 
%One of the fascinating property of a single 
%isotropic chain is that their energy spectrum is gapless for half-integer spin length 
%and gapped otherwise. Due to the proportionality between the entanglement spectrum of 
%the ladders and energy spectrum of the single chain, the entanglement spectrum of the 
%ladder has the same property, even the energy spectrum is always 
%gapped for any spin length in case of large rung coupling. The inverse temperature
%is independent of a spin length. 
However, they found that there is no proportionality 
between the entanglement Hamiltonian of anisotropic spin ladders of arbitrary
spin length. Since here, the reduced density matrix in zeroth order of the perturbation
theory is not proportional to the unit matrix, there is no mention of proportionality.

\section{Model}
\label{model}
We investigate the Hamiltonian of the Heisenberg spin-1/2 ladder in a time-dependent 
circularly polarized magnetic field $ B $ described by the Hamiltonian
\begin{align}
\tilde{H}&=J_{rung}\sum_{i}\vec{S}_{2i}\vec{S}_{2i+1} 
+B\sum_{i}\left(S_{i}^{x}\cos\omega t-S_{i}^{y}\sin\omega t\right) \nonumber \\
&+J_{leg}\sum_{i}\vec{S}_{2i}\vec{S}_{2i+2}
+J_{leg}\sum_{i}\vec{S}_{2i+1}\vec{S}_{2i+3}. \label{Ham_cir}
\end{align} 
where $ \omega $ is the angular velocity of the rotation of the magnetic field. 
The sites on the first (second) leg are denoted by even (odd) labels, such 
that the $ i $th rung consists of sites $ 2i $ and $ 2i+1 $. 
All spin-1/2 operators are taken to be dimensionless, such that the couplings along 
the rungs $ J_{rung} $ and the legs $ J_{leg} $ have the dimensions of energy. 
We will consider antiferromagnetic coupling when $ J_{rung}>0 $.

This time-dependent Hamiltonian can be factorized to a time-independent Hamiltonian 
by unitary transformations that represent a rotation around the z-axis 
$ R(t)=\exp(-iS_{z}\omega t/ \hbar) $ \cite{QT}.
Since
\begin{align}
&R(t)S_{x}R^{-1}(t)=S_{x}\cos \omega t+S_{y}\sin \omega t \nonumber \\
&R(t)S_{y}R^{-1}(t)=-S_{x}\sin \omega t+S_{y}\cos \omega t \nonumber \\
&R(t)S_{z}R^{-1}(t)=S_{z} \label{transformations}
\end{align} 
Hamiltonian Eq. (\ref{Ham_cir}) can be transformed into a time-independent Hamiltonian
\begin{align}
\widehat{H}&=R(t)\tilde{H}(t)R(t)^{-1}\\
\widehat{H}&=J_{rung}\sum_{i}\vec{S}_{2i}\vec{S}_{2i+1}+B\sum_{i}S_{i}^{x} 
+J_{leg}\sum_{i}\vec{S}_{2i}\vec{S}_{2i+2} \nonumber \\
&+J_{leg}\sum_{i}\vec{S}_{2i+1}\vec{S}_{2i+3}. \label{Ham_tid}
\end{align}

Defining the propagator that confirms
\begin{align}
&\frac{\partial}{\partial t}K(t,t_{0})=-\frac{i}{\hbar}\tilde{H}(t)K(t,t_{0}) \\
&\frac{\partial}{\partial t}K(t,t_{0})=-\frac{i}{\hbar}R_{-1}(t)\widehat{H}(t)R(t)K(t,t_{0})
\end{align}
we find
\begin{eqnarray}
\frac{\partial}{\partial t}\left(R(t)K(t,t_{0})R^{-1}(t_{0})\right)=-\frac{i}{\hbar}H\left(R(t)K(t,t_{0})R^{-1}(t_{0})\right). \nonumber
\end{eqnarray} 
Then, the Hamiltonian becomes
\begin{align}
H=&J_{rung}\sum_{i}\vec{S}_{2i}\vec{S}_{2i+1}
+B\sum_{i}S_{i}^{x}+\omega\sum_{i}S_{i}^{z} \nonumber \\
+&J_{leg}\sum_{i}\vec{S}_{2i}\vec{S}_{2i+2} 
+J_{leg}\sum_{i}\vec{S}_{2i+1}\vec{S}_{2i+3}\label{Hami}
\end{align}
where the propagator is
\begin{footnotesize}
\begin{eqnarray}
K(t,t_{0})=\exp\left(\frac{i}{\hbar}S_{z}\omega t\right)
\exp\left(-\frac{i}{\hbar}H(t-t_{0})\right)\exp
\left(-\frac{i}{\hbar}S_{z}\omega t_{0}\right).
\label{propagator}
\end{eqnarray}
\end{footnotesize}
In order to use the perturbation theory, we will rewrite the Hamiltonian 
Eq.(\ref{Hami}) as $ H=H_{0}+H_{1} $ where
\begin{eqnarray}
H_{0}=J_{rung}\sum_{i}\vec{S}_{2i}\vec{S}_{2i+1}+B\sum_{i}S_{i}^{x}+\omega\sum_{i}S_{i}^{z}  
\label{Hami0}
\end{eqnarray}
and
\begin{eqnarray}
H_{1}=J_{leg}\sum_{i}\vec{S}_{2i}\vec{S}_{2i+2}
+J_{leg}\sum_{i}\vec{S}_{2i+1}\vec{S}_{2i+3} \label{Hami1}
\end{eqnarray}
and consider $ H_{1} $ as a small perturbation.
The Hamiltonian Eq.(\ref{Hami0}) is independent of the 
direction of the magnetic field and 
it can be considered as the isotropic Heisenberg chain in the magnetic field 
$ \sqrt{B^{2}+\omega^{2}} $
\begin{eqnarray}
H_{0}&=J_{rung}\sum_{i}\vec{S}_{2i}\vec{S}_{2i+1}
+\sqrt{B^{2}+\omega^{2}}\sum_{i}S_{i}^{z} \label{Hami10}
\end{eqnarray}
and
\begin{eqnarray}
H_{1}&=J_{leg}\sum_{i}\vec{S}_{2i}\vec{S}_{2i+2}
+J_{leg}\sum_{i}\vec{S}_{2i+1}\vec{S}_{2i+3}. 
\label{Hami11}
\end{eqnarray}

The energies of a rung of the singlet and triplet states are
\begin{align}
&E_{s_{i}}=-\frac{3}{4} J_{rung}\\
&E_{t_{i}^{+}}=\frac{1}{4} J_{rung}+\sqrt{B^{2}+\omega^{2}} \\
&E_{t_{i}^{0}}=\frac{1}{4} J_{rung} \\
&E_{t_{i}^{-}}=\frac{1}{4} J_{rung}-\sqrt{B^{2}+\omega^{2}},
\end{align}

The ground state changes from the spin singlet $ \vert s_{i} \rangle $,
to the triplet state $ \vert t^{-}_{i} \rangle $
by increasing the value of $ \sqrt{B^{2}+\omega^{2}} $ .
When $ J_{rung}= \sqrt{\omega^{2}+B^{2}} $, the ground state 
is two-fold degenerate, since the singlet states $ \vert s_{i}\rangle $ 
and triplet states $ \vert t_{i}^{-}\rangle $ have the same eigenenergy.
The situation when ground state is two-fold degenerate is quite interesting
and it will be considered in the following section.

\section{Entanglement spectrum}
When $ J_{rung}= \sqrt{\omega^{2}+B^{2}} $, 
it is necessary to use degenerate perturbation theory, while
any combination of eigenstates $ \vert s_{i}\rangle $ 
and $ \vert t_{i}^{-}\rangle $ can be taken as the ground state 
$ \vert \psi_{0} \rangle $.
In order to achieve an analytically manageable situation, we will assume
a finite number of rungs $ i = 4 $. 
Let us suppose that the unperturbed ground state $ \vert psi_0\rangle $ 
of the Hamiltonian $ \mathcal{H}_{0} $ is an unknown 
combination of eigenvectors $ \vert s_{i}\rangle $
and $ \vert t_{i}^{-}\rangle $. In the following, we will note 
eigenvectors of the ground state $ \vert \psi_0\rangle $ as
$ \lbrace\vert \overline{n}\lambda \rangle \vert, \lambda=1,...,4\rbrace $,
where 
\begin{align}
&\vert\overline{n}1\rangle=\vert s_{1}\rangle \vert s_{2}\rangle 
\vert s_{3}\rangle \vert s_{4}\rangle ,
 \vert\overline{n}2\rangle=\vert s_{1}\rangle \vert s_{2}\rangle 
\vert s_{3}\rangle \vert t_{4}^{-}\rangle, \nonumber \\
&\vert\overline{n}3\rangle=\vert s_{1}\rangle \vert s_{2}\rangle 
\vert t_{3}^{-}\rangle \vert s_{4}\rangle, 
\vert\overline{n}4\rangle=\vert s_{1}\rangle \vert t_{2}^{-}\rangle 
\vert s_{3}\rangle \vert s_{4}\rangle,  \nonumber \\
&\vert\overline{n}5\rangle=\vert t_{1}^{-}\rangle \vert s_{2}\rangle 
\vert s_{3}\rangle \vert s_{4}\rangle, 
\vert\overline{n}6\rangle=\vert t_{1}^{-}\rangle \vert t_{2}^{-1}\rangle 
\vert s_{3}\rangle \vert s_{4}\rangle, \nonumber \\
&\vert\overline{n}7\rangle=\vert t_{1}^{-1}\rangle \vert s_{2}\rangle 
\vert t_{3}^{-1}\rangle \vert s_{4}\rangle, 
\vert\overline{n}8\rangle=\vert t_{1}^{-1}\rangle \vert s_{2}\rangle 
\vert s_{3}\rangle \vert t_{4}^{-1}\rangle, \nonumber \\
&\vert\overline{n}9\rangle=\vert s_{1}\rangle \vert t_{2}^{-1}\rangle 
\vert t_{3}^{-1}\rangle \vert s_{4}\rangle,  
\vert\overline{n}10\rangle=\vert s_{1}\rangle \vert t_{2}^{-1}\rangle 
\vert s_{3}\rangle \vert t_{4}^{-1}\rangle, \nonumber \\
&\vert\overline{n}11\rangle=\vert s_{1}\rangle \vert s_{2}\rangle 
\vert t_{3}^{-1}\rangle \vert t_{4}^{-1}\rangle, 
\vert\overline{n}12\rangle=\vert t_{1}^{-1}\rangle \vert t_{2}^{-1}\rangle 
\vert t_{3}^{-1}\rangle \vert s_{4}\rangle, \nonumber \\
&\vert\overline{n}13\rangle=\vert t_{1}^{-1}\rangle \vert t_{2}^{-1}\rangle 
\vert s_{3}\rangle \vert t_{4}^{-1}\rangle,
\vert\overline{n}14\rangle=\vert t_{1}^{-1}\rangle \vert s_{2}\rangle 
\vert t_{3}^{-1}\rangle \vert t_{4}^{-1}\rangle, \nonumber \\
&\vert\overline{n}15\rangle=\vert s_{1} \rangle \vert t_{2}^{-1}\rangle 
\vert t_{3}^{-1}\rangle \vert t_{4}^{-1}\rangle, 
\vert\overline{n}16\rangle=\vert t_{1}^{-1}\rangle \vert t_{2}^{-1}\rangle 
\vert t_{3}^{-1}\rangle \vert t_{4}^{-1}\rangle
\end{align}
The projector $ P_{n}^{(0)} $ of $ \mathcal{H}_{0} $ projects
on the subspace, and is defined by the eigenvalue $ E_{\overline{n}}^{(0)}=-\frac{3}{4} J_{rung} $
of the Hamiltonian $ \mathcal{H}_{0} $.
Furthermore, the projector $ P_{n}^{(0)} $ satisfies
\begin{eqnarray}
P_{n}^{(0)} \mathcal{H}^{'} P_{n}^{(0)} \vert \psi_{0} \rangle =
E^{(1)} \vert \psi_{0} \rangle
\end{eqnarray}
where $ E^{(1)} $ is the eigenvalue of $ P_{n}^{(0)} \mathcal{H}^{'} P_{n}^{(0)} $
for eigenvector $ \vert \psi_{0} \rangle $. In order to find the eigenvalue 
$ E^{(1)} $ of the perturbation $ \mathcal{H^{'}} $ and the ground state 
$ \vert \psi_0 \rangle $, it is sufficient to diagonalize a $16\times16$ matrix
\begin{equation}
\left[
\begin{array}{ccc}
\langle \overline{n} 1 \vert \mathcal{H}^{'} \vert \overline{n} 1 \rangle & 
\cdot \cdot \cdot  & 
\langle \overline{n} 1 \vert \mathcal{H}^{'} \vert \overline{n} 16 \rangle \\
\cdot \cdot \cdot  & \cdot \cdot \cdot & \cdot \cdot \cdot \\ 
\langle \overline{n} 16 \vert \mathcal{H}^{'} \vert \overline{n} 1 \rangle &
\cdot \cdot \cdot & \langle \overline{n} 16 \vert \mathcal{H}^{'} \vert \overline{n} 16 \rangle 
\end{array}
\right].
\end{equation}
By elementary calculations, one finds the uniquely determined ground state $ 
\vert \psi_0 \rangle $ and the first correction of the energy $ E^{(1)} $.

The unperturbed density matrix is constructed from this ground state and is given,
after simplification, by
\begin{eqnarray}
\rho^{(0)}=& \sum_{\lambda}^{16} \vert \overline{n} \lambda \rangle \langle 
\overline{n} \lambda \vert.
\end{eqnarray}
By again tracing out one leg, we obtain the reduced unperturbed density matrix
\begin{eqnarray}
\rho_{red}^{(0)}=\frac{1}{2^{4}}\bigotimes_{i=1}^{4}\left(1-S_{2i+1}^{z}\right)
\end{eqnarray}
It is obvious that this reduced density matrix is not proportional to the unitary matrix
and possesses a nontrivial entanglement spectrum.
The first corrections to the ground state in the degenerate perturbation theory
are defined by
\begin{align}
\vert 1\rangle=&\sum_{n\neq \overline{n}} \vert n \lambda \rangle 
\frac{\langle n \lambda \vert H_{1} \vert \psi_{0}\rangle}
{E_{\overline{n}}^{(0)}-E_{n}^{(0)}}, \\
\vert 1^{'}\rangle=&\sum_{n\neq \overline{n},n'} \sum_{\lambda'} 
\vert \overline{n} \lambda \rangle 
\frac{\langle \overline{n} \lambda \vert H_{1} \vert n' \lambda\rangle}
{E^{(1)}-E_{\lambda}^{(1)}}
\frac{\langle n' \lambda \vert H_{1} \vert \psi_{0}\rangle}
{E_{\overline{n}}^{(0)}-E_{n'}^{(0)}}.
\end{align}
One finds the reduced density matrix to the first order
\begin{align}
\rho_{red}^{(1)}=&\frac{1}{2^{4}}\bigotimes_{i=1}^{4}\left(1-S_{2i+1}^{z}\right) 
-\frac{J_{leg}}{8J_{rung}}\left((1-S_{1}^{z})(1-S_{3}^{z})\vec{S}_{5}\vec{S}_{7}\right.
\nonumber \\
+&\left.(1-S_{1}^{z})\vec{S}_{3}\vec{S}_{5}(1-S_{7}^{z}) 
+\vec{S}_{1}\vec{S}_{3}(1-S_{5}^{z})(1-S_{7}^{z})
\right)
\label{rdm}
\end{align}

The reduced density matrix can be rewritten as
\begin{eqnarray}
\rho_{red}=\frac{1}{Z}\exp(-\mathcal{H}_{ent}^{(1)})
\end{eqnarray}
where the partition function $Z$ is $ Z=\textrm{tr} \exp(-\mathcal{H}_{ent}^{(1)}) $
and the entanglement Hamiltonian within the first order of the perturbation theory 
has the following form
\begin{eqnarray}
\mathcal{H}_{ent}^{(1)}=\frac{1}{J_{rung}}\sum_{i=0}^{3}\left(2J_{leg}\vec{S}_{2i+1}\vec{S}_{2i+3}+
J_{rung} S_{2i+1}\right).
\label{ent_Ham1}
\end{eqnarray}
The entanglement Hamiltonian is simply proportional to the 
Hamiltonian of a chain in the magnetic field with the proportional factor 
$ \beta=\frac{1}{J_{rung}} $ defined as an inverse temperature.

The system of the Heisenberg chain in the longitudinal magnetic field is 
exactly solvable by the Bethe ansatz. The ground state becomes the spin--liquid
one and gapless up to when $ \frac{\sqrt{B^{2}+\omega^{2}}}{J_{leg}} = 2 $, 
where the phase transition of the Pokrovsky--Talapov type takes place and the 
ground state becomes a completely ordered gapped ferromagnetic state.
One of the most important features of the energy spectra of spin chains is the absence 
of an excitation gap over the ground state for the integer spin length.
We restrict ourselves to the case when the two chains are strongly 
coupled; therefore, $ \frac{\sqrt{B^{2}+\omega^{2}}}{J_{leg}} >> 1 $ and the Hamiltonian
of a subsystem stays gapless in this region. The entanglement spectrum 
Eq.(\ref{ent_Ham1}) remains gapless owing to the proportionality to the energy spectrum.

\section{Summary}
\label{conclusion2}
Here we investigated the entanglement spectrum of Heisenberg ladders in a 
time-dependent magnetic field using the degenerate perturbation theory, where couplings
between legs are taken as a small perturbation. When the ground state is not
degenerate, the existence of the trivial entanglement Hamiltonian in the zeroth order of the perturbation
theory is identified as an important condition that guarantees the proportionality
between the entanglement and subsystem Hamiltonians. 
We find that although the entanglement spectrum of the unperturbed
ground state has a nontrivial entanglement spectrum,
the entanglement Hamiltonian of Heisenberg ladders in a 
time-dependent magnetic field, within the first-order perturbation theory,
is proportional to the energy Hamiltonian of a chain in the magnetic field
when the ground state is degenerate.

\section{ACKNOWLEDGMENTS}

The author kindly acknowledges John Schliemann.

{}

\end{document}